\documentclass[a4paper]{llncs}
\usepackage{url}
\usepackage{graphicx}
\usepackage{makeidx}  
\usepackage{color}
\begin{document}

\mainmatter  

\title{Development of an Ontology for an Integrated Image Analysis Platform to enable Global Sharing of Microscopy Imaging Data}
\author{Satoshi Kume\inst{1,2} \and Hiroshi Masuya\inst{3,4} \and Yosky Kataoka\inst{1,2} \and Norio Kobayashi\inst{2,3,4}\thanks{To whom correspondence should be addressed: norio.kobayashi@riken.jp}}

\institute{
RIKEN Center for Life Science Technologies (CLST),\\
6-7-3 Minatojima-minamimachi, Chuo-ku, Kobe, Hyogo, 650-0047 Japan\\
\{satoshi.kume, kataokay\}@riken.jp
\and
RIKEN CLST-JEOL Collaboration Center, \\
6-7-3 Minatojima-minamimachi, Chuo-ku, Kobe, Hyogo, 650-0047 Japan\\
\and
RIKEN BioResource Center (BRC),\\
3-1-1, Koyadai,Tsukuba, Ibaraki, 305-0074 Japan\\
hmasuya@brc.riken.jp
\and
Advanced Center for Computing and Communication (ACCC), RIKEN,\\
2-1 Hirosawa, Wako, Saitama, 351-0198 Japan\\
norio.kobayashi@riken.jp
}

\maketitle 

\begin{abstract}
Imaging data is one of the most important fundamentals in the current life sciences. 
We aimed to construct an ontology to describe imaging metadata
as a data schema of the integrated database for optical and electron microscopy images combined with various bio-entities. To realise this, we applied Resource Description Framework (RDF) to an Open Microscopy Environment (OME) data model, which is the de facto standard to describe optical microscopy images and experimental data. We translated the XML-based OME metadata into the base concept of RDF schema 
as a trial of developing microscopy ontology. In this ontology, we propose
18 upper-level concepts including missing concepts in OME such as electron microscopy, phenotype data, biosample, and imaging conditions.
\end{abstract}

\setcounter{footnote}{0}

\section{Introduction}
Imaging data is a crucial fundamental in current life science. Recently, ultra-microstructural imaging data, such as scanning electron microscopy (SEM), has provided detailed morphological phenomena of tissues and/or cells \cite{Kas2015}: moreover, SEM imaging analysis, such as image segmentation and reconstruction, is a very interesting topic for big data  and computational science. We are working on the production of the large-scale nano-scale microstructural imaging data of mammalian (mouse and rat) tissues using SEM 
and plan to develop open-accessible metadatabase of microstructural imaging data from organelles in cells to the tissue structure (\url{http://www.clst.riken.jp/en/science/labs/collabo/}).

To share microscopy images and experimental data, we have employed the Resource Description Framework (RDF), which is a standardised framework for data sharing to enable interoperable metadata integration and multidisciplinary data. However, standardised ontologies that describe optical and electron microscopy (EM) metadata, including biosamples, bio-resources, experimental conditions, have not been developed, and integrated analysis of imaging data with other metadata remains difficult.

The Open Microscopy Environment (OME), an open-source interoperability toolset for biological imaging data, has been proposed for managing multidimensional and heterogeneous imaging data for optical microscopy \cite{Allan2012}. 
The translation of OME data model in XML to OWL/RDF was previously performed \cite{RBE2004}. However, in the old version, extension toward data integration with biosamples, experimental conditions and other microscopy modalities such as EM were not well examined. 
In this paper, we tried to construct RDF-based data model of latest version of OME 
(\url{https://github.com/kumeS/ISWC2016}) for standardised description of microscopy image data.

\section{Method}
\subsubsection{Translation of OME data model to RDF}
We extracted key concepts and properties from the OME data model (version: January 2015) written in XML schema. Then, we reconstructed the relations between RDF properties and XML elements using the extracted OME properties.

\subsubsection{Extension of OME data model with biosamples and electron microscopy images in RDF}
To construct a data-interoperable ontology with existing RDF-based biological and medical data and concepts, we expanded upper-level concepts and properties extracted from above RDF (e.g. electron microscopy devices and bio-resources). In particular, we have added new vocabularies for EM experiments, including phenotype data, biosamples and bio-resources and experimental conditions such as imaging conditions and sample preparations, as an extension of the RDF version of the OME data model. 

\section{Results and Discussion}
\begin{figure}[ht]
\centering
\vspace*{-2mm}
\includegraphics[width=11cm]{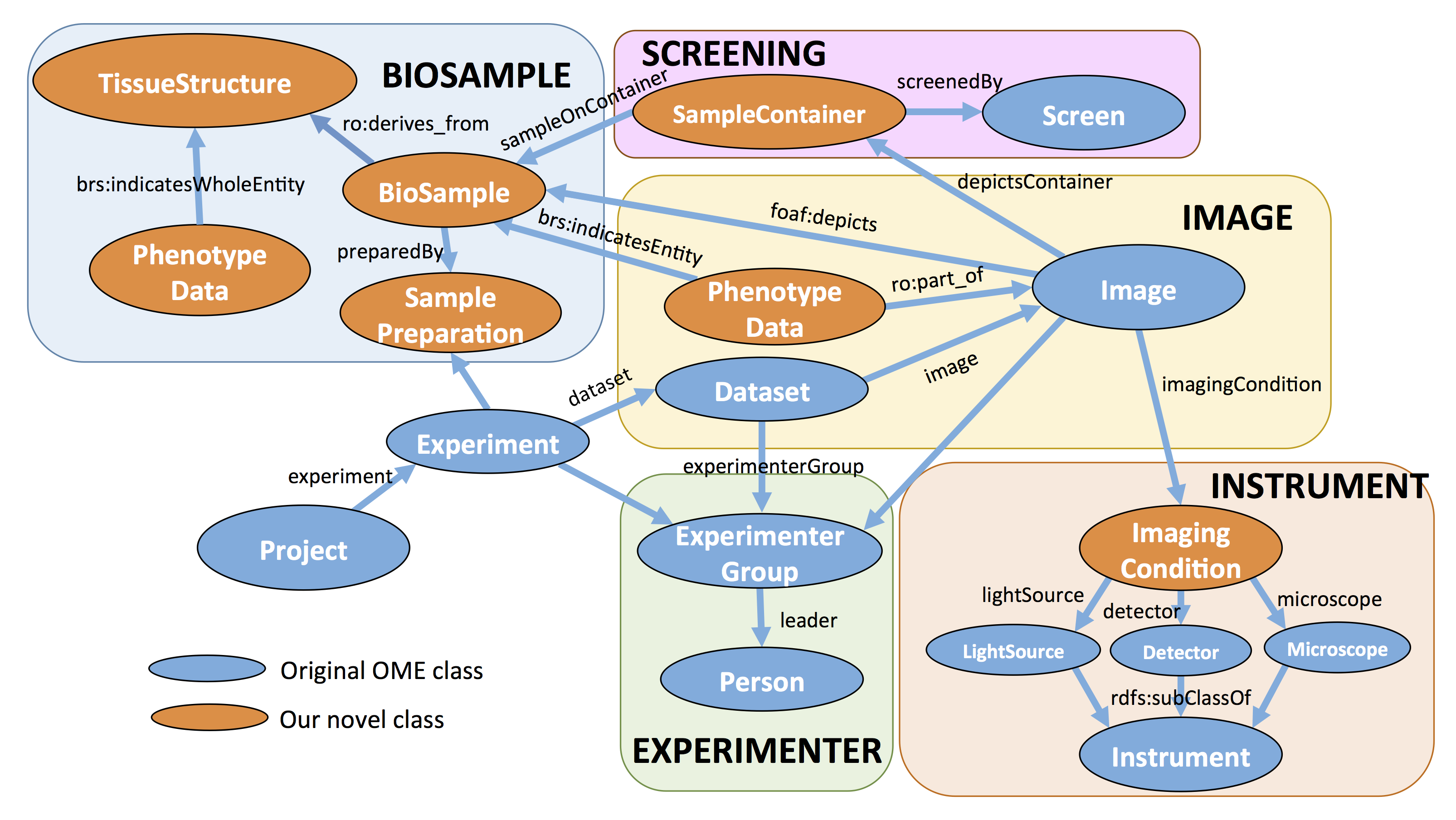}
\caption{Overview of classes and relations extracted from the OME data model. The blue classes are directly translated classes from the OME data model, and the orange classes are the proposed extended classes.}
\label{fig:overview}
\end{figure}

Referencing of image data from experimental results has become more important in current life science. Thus, we aimed to expand global data integration by the construction of RDF schema describing imaging metadata in various biological analyses as well as electron microscopy.
The upper-level concepts and the relation graph of RDF-based data model was constructed from current version of OME and the schema is summarized in Fig. \ref{fig:overview}. The proposed RDF/OWL comprises five categories, i.e. IMAGE, EXPERIMENTER, INSTRUMENT, BIOSAMPLE and SCREENING. 
It also contains 18 upper-level ontology concepts, including Image, SampleContainer, BioSample, PhenotypeData and ImagingCondition classes, which represent the 7 concepts shown in orange in Fig. \ref{fig:overview}.

\begin{figure}[ht]
\centering
\vspace*{-3mm}
\includegraphics[width=11cm]{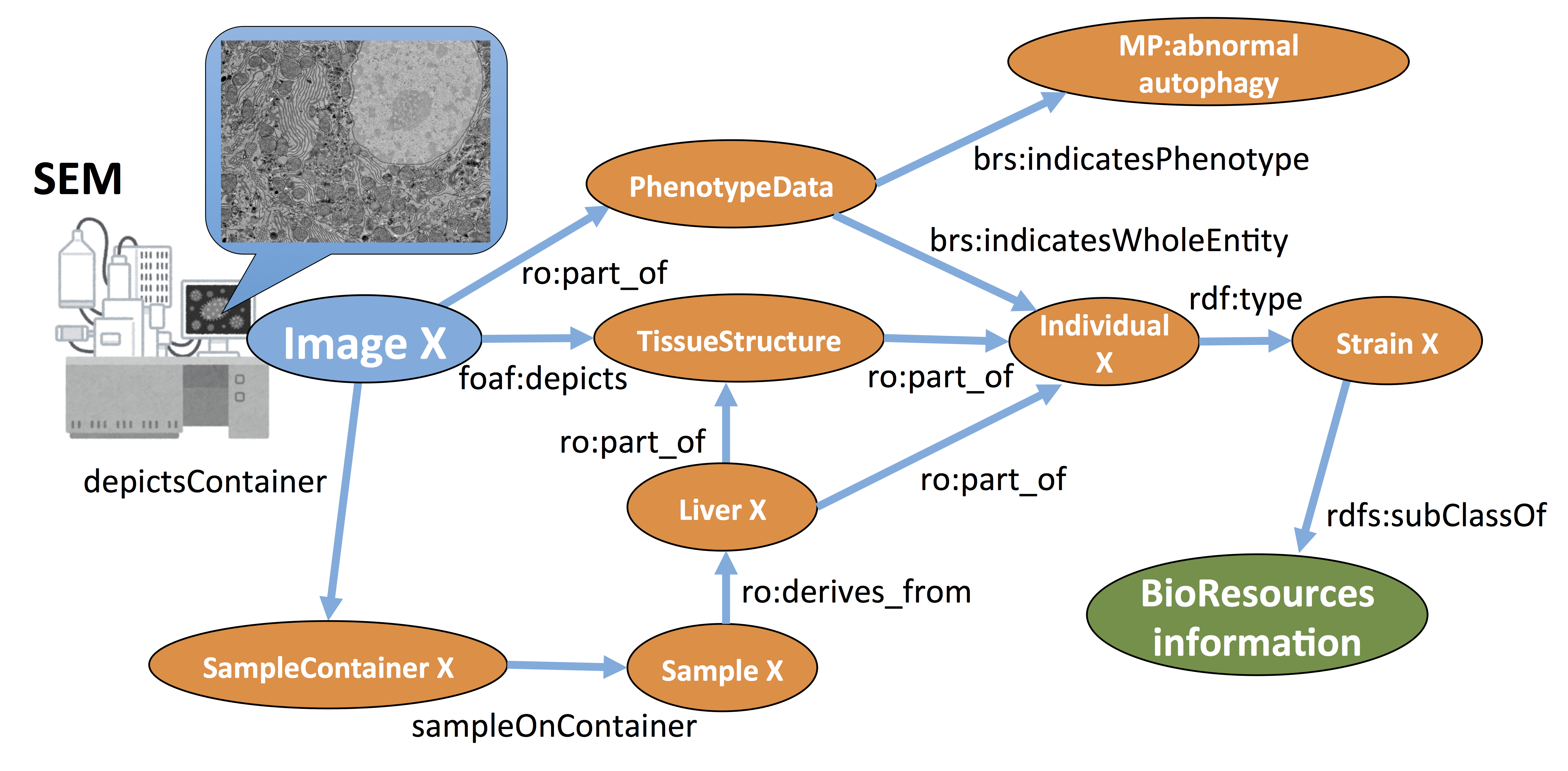}  
\caption{Extended graph structure that describes images with detailed biosample information used in an experiment. The blue ovals are instances of classes translated from the OME data model and the orange ovals are instances of classes as an extension represented in Fig. 1. The green oval represents bioresource information as a link to an external database.}
\label{fig:classRel}
\end{figure}

The OME data model primarily focused on optical microscopy devices and image data. 
To describe EM devices and imaging conditions, we included class descriptions such as electron wave and electron gun, in the concept of imaging conditions. 
Furthermore, to integrate the microscopy metadata and other bioresource databases, we added "Biosample" and "Bioresource" classes. 
An example of an RDF graph written with our ontology is shown in Fig. \ref{fig:classRel}.
In this example, the entities that we want to observe (sample container and sample) and observation results (liver cells, tissue structure and phenotype data) can be precisely described. Through the sample and strain RDF entities, microscopy images can be linked to detailed information described by another database and ontology, such as the RIKEN BioResource Center (\url{http://metadb.riken.jp/metadb/db/rikenbrc_mouse}).
Using the ontology, we tried to describe about 20,000 imaging data from EM which are obtained from different staining methodologies in the sample preparation. As a result, we successfully described these data sets (data not shown).

%
%
%
%
%
%

In conclusion, our microscopy ontology based on the OME data model can be applicable to imaging data and associated information obtained by optical and electron microscopy devices and can reorganise the image data for integrated image analysis. In future, we will expand the ontology to cooperate with Biological Dynamics Markup Language \cite{Kyoda2014} and Cellular Microscopy Phenotype Ontology \cite{Jupp2016}. We also plan to generate electron microscopy images metadata in RDF through the RIKEN MetaDatabse (\url{http://metadb.riken.jp}) platform which enables the integration with RIKEN's various life-science data and globally published linked open data. 

\subsubsection*{Acknowledgments} 
This work was supported by the Management Expenses Grant for RIKEN CLST-JEOL Collaboration Center.


%
%

\end{document}